\documentclass[preprint,showpacs,preprintnumbers,amsmath,amssymb,floatfix]{revtex4}
\usepackage[dvips]{graphicx}
\usepackage{amsmath}
\usepackage{bm}
\usepackage[dvips]{graphicx}
\usepackage{amsfonts}
\usepackage{bm}
\begin{document}

\title{Collapse of Telechelic Star Polymers
to Water-Melon Structures}
\author{Federica Lo Verso}
\author{Christos N. Likos}
\author{Christian Mayer}
\author{Hartmut L\"owen}
 \affiliation{Institut f\"ur Theoretische Physik II,
Heinrich-Heine-Universit\"at D\"usseldorf, 
D-40225 D\"usseldorf, Germany}
\date{\today}

\begin{abstract}
Conformational properties of star-shaped polymer aggregates that 
carry attractive
end-groups, called {\it telechelic star polymers}, are investigated by 
simulation and analytical variational theory. We focus
on the case of low telechelic star polymer functionalities, $f \leq 5$,
a condition which allows aggregation 
of all attractive monomers on one site.
We establish the functionality- and
polymerization-number dependence of the transition temperature 
from the ``star burst'' to the
``water melon" macroparticle structure. Extensions to 
telechelic stars featuring partially collapsed configurations
are also discussed.
\end{abstract}

\maketitle

%
%
Self-organizing soft materials are relevant in developing 
novel macromolecular compounds with peculiar structural and 
dynamical properties, which are related to mesoscopic aggregation 
as well as intra- and inter-molecular association 
\cite{muthu:science:97,mcleish:book}.
Progresses in the synthesis of chains with
attractive end-groups, called {\it functionalized} or
{\it telechelic polymers}, 
opened the way for the subsequent synthesis of telechelic star polymers
with attractive polar end groups, telechelic associating polymers with 
hydrophobic terminal groups, 
associating polyelectrolytes in homogeneous solutions,
and telechelic planar brushes 
\cite{pitsikalis:macrom:96,vlassopoulos:jcp:99,clement:macrom:00,zaroslov:05}.
A telechelic star, which is the subject of the work at hand, consists of
$f$ chains with one of their ends chemically attached on a common center,
whereas the attractive ends are free at the other end of the molecule.
Block copolymer stars with a hydrophilic core and a hydrophobic 
corona \cite{ganazzoli:mts:01,connolly:jcp:03}
also bear similarities to telechelic stars, the two systems becoming
identical when the length of the hydrophobic block reduces to a single
monomer.
The thermodynamics and the structure of planar telechelic brushes have been
recently analyzed also from a theoretical point of view, 
shedding light into the quantitative characteristics of both the
conformations and the interactions of the same 
\cite{semenov:macrom:95,zilman:epje:01,russel:macrom:03}.
Further theoretical approaches have been developed to
describe flower-like micelles 
with hydrophobic terminal groups that self-assemble in water.
Such aggregates show a characteristic `bridging attraction' 
\cite{semenov:macrom:95,russel:macrom:00} that can
even lead to a liquid-vapor phase transition \cite{russel:macrom:97}.
The interesting feature of 
telechelic star polymers 
is the possibility of attachment between
the attractive terminal groups.
In dilute solutions, this gives rise to intra-molecular association 
as well as inter-association between micelles, 
depending on the details of the molecular structure. 

Examples of
low-functionality telechelics, very similar to the ones considered in
this work, are mono-, di-, and
tri-$\omega$-zwitterionic, three-arm star symmetric
polybutadienes.
Experiments 
have shown that
these self-assemble into distinct supramolecular structures, including
collapsed, soft-sphere
conformations \cite{vlassopoulos:jcp:99,vlassopoulos:macrom:00}.
In particular, using low-angle laser light scattering
and dynamic light scattering, 
it was found that samples with three zwitterion end groups 
present a low degree of inter-association between macromolecules,
showing instead a preference for intra-association and formation
of collapsed soft spheres \cite{pitsikalis:macrom:96,pitsikalis:macrom:95}.
X-ray scattering and rheological experiments support the conclusion that
this tendency persists at higher concentrations, all the 
way into the melt \cite{vlassopoulos:jcp:99}. There, the formation
of transient gels has been found for the case of two- and
three-zwitterion macromolecules, with the network characteristics
depending on the molecular weight of the arms \cite{vlassopoulos:macrom:00}.

The conformations of telechelic micelles are evidently determined 
by the 
competition between entropic and energetic contributions. 
A detailed investigation, by theory and simulation, 
of the mechanisms leading to the formation of collapsed soft spheres
is, however, still lacking.
In this Letter we perform extensive computer simulations,
accompanied by a scaling analysis of the free energies of candidate
structures of 
telechelic micelles with small functionality, $f\leq 5$. We find that
at high temperatures the system 
assumes the usual star burst ({\it sb}) configuration.
On the contrary, at sufficiently low temperatures, 
the end-monomers attach to each other and the micelles assume an
overall closed configuration, akin to the collapsed soft spheres
conjectured in the experimental study of Ref.\ \cite{vlassopoulos:jcp:99}.
Due to the peculiar shape of these aggregates, featuring two points
of aggregation, one at each end, we term them ``water melons'' ({\it wm}). 
For intermediate temperatures, 
the macromolecules exhibit several configurations that 
correspond  to a partial assembling of the terminal groups.
The number of chains connected 
at their ends depends on the temperature $T$,
the functionality $f$ and the degree of polymerization $N$ 
of the chains.
In particular, at a given, sufficiently low $T$, the {\it wm}-configuration
becomes more stable with respect to the {\it sb}-one with decreasing
$N$ and increasing $f$, as will be demonstrated in what follows.
Our theoretical predictions are found to agree well with simulation 
results.

%
%
We employed monomer-resolved molecular dynamics 
(MD) simulations to examine 
the conformations
of isolated telechelic micelles. The monomers were modeled as soft
spheres interacting by means of a truncated and shifted 
Lennard-Jones potential, $V_{\rm LJ}^{\rm tr}(r)$ (length scale
$\sigma_{\rm LJ}$, energy scale $\varepsilon$), the truncation
point being at the minimum, $r_{\rm min} = 2^{1/6}\sigma_{\rm LJ}$,
rendering the interaction purely repulsive. The {\it end-monomers},
on the other hand, interact with each other by means of the {\it full}
 Lennard-Jones potential, $V_{\rm LJ}(r)$ that has a minimum 
value $V_{\rm LJ}(r_{\rm min}) = -\varepsilon$. The chain connectivity
was modeled by employing the finite extension nonlinear elastic 
(FENE) potential \cite{fene:93}, using the same parameter
values as in the simulation of normal star polymers 
\cite{grest:macrom:87, grest:review}. 
The monomer mass $m$, the energy $\varepsilon$ and the length
$\sigma_{\rm LJ}$ are assigned the value unity, defining thereby the
characteristic time $\tau = m\sigma_{\rm LJ}^2/\varepsilon$ and the
dimensionless temperature $T^* = k_{\rm B}T/\varepsilon$, where
$k_{\rm B}$ is Boltzmann's constant. 
The equations of motion were integrated using the velocity form of 
Verlet's algorithm \cite{allen:tildesley}, whereas the temperature
was fixed by applying 
a Langevin thermostat \cite{langevin}. The timestep was
$\Delta t = 10^{-3}\tau$, with a total of $2\times 10^5$ timesteps
used for equilibration and 10 independent runs 
of $5\times 10^7$ timesteps each to gather statistics.
The characteristic quantities measured were
the radius of gyration $R_g$ of the molecules, the radial 
distribution function $g_t(r)$ between the terminal monomers
and the expectation value $E_t$ of the interaction
between the terminal groups, with
$E_t = \langle \sum_{i=1}^f\sum_{j > i} V_{\rm LJ}(|{\bf t}_i - {\bf t}_j|)
\rangle$, where $\langle\cdots\rangle$ denotes a statistical 
average and ${\bf t}_{i}$ stands for the position vector of the
end-monomer of the $i$-th chain.

\begin{figure}
\begin{center}
\includegraphics[width=14.cm,angle=0.,clip]{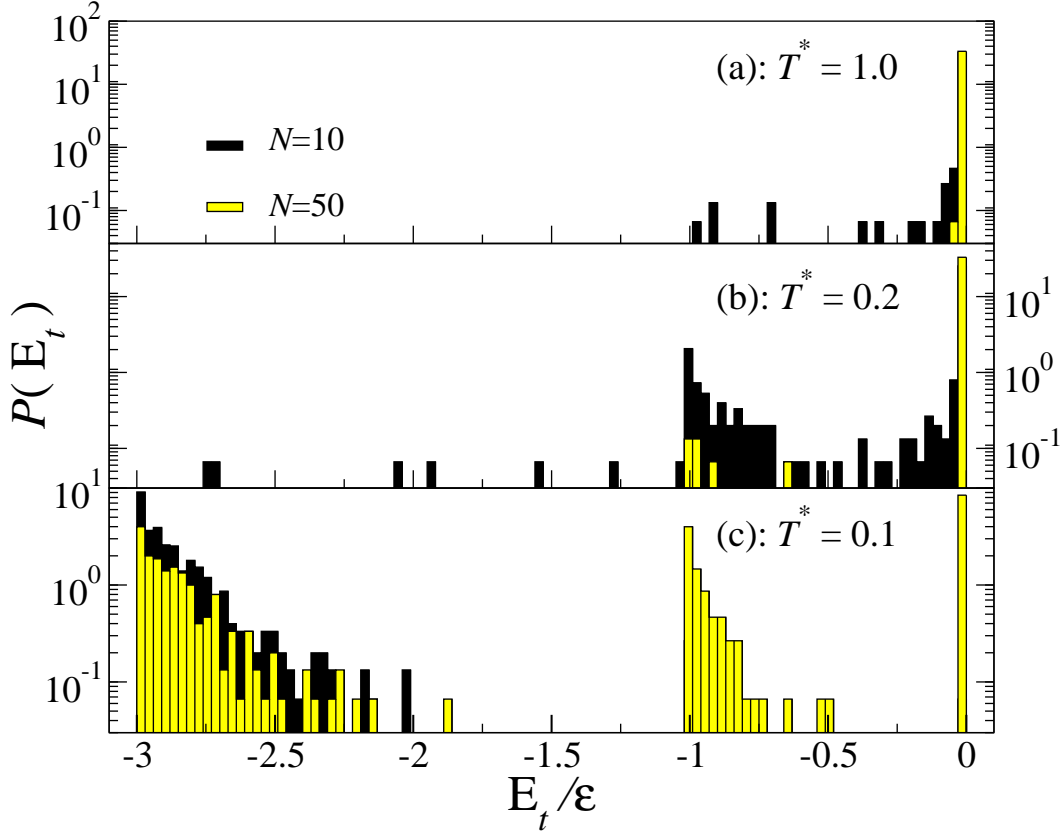}
\caption{The probability distribution function $P(E_t)$
obtained from simulations of $f=3$ telechelic micelles carrying
$N=10,50$ monomers per arm. Note the logarithmic scale.}
\label{fig1:plot}
\end{center}
\end{figure}
We considered temperatures between $T^*=0.01$ and 
$T^*=1.2$.
For $T^* \gtrsim 0.2$, we found that the micelle configuration is similar 
to that of a star polymer without attractive ends, i.e., the
{\it sb}-configuration. 
A morphological transition of the micelles takes place
upon lowering the temperature. 
In Fig.\ \ref{fig1:plot} we show the probability distribution
function $P(E_t)$ for the end-monomer interaction
energy $E_t$ for the case $f=3$, for different $T^{*}$- and $N$-values.
Let us focus on $N=10$.
At $T^* = 1.0$, Fig.\ \ref{fig1:plot}(a),
$P(E_t)$ shows 
almost all the events around
$E_t = 0$, meaning that in the vast majority of 
configurations the end-monomers are far apart, so
that the attractions are vanishing. 
For $T^*=0.2$, Fig.\ \ref{fig1:plot}(b), 
$P(E_t)$ takes a bimodal form
with two peaks, one at $E_t = 0$ and one at
$E_t = -\varepsilon$. The latter corresponds to a 
conformation in which two chains are end-attached, with their
terminal monomers at a distance $r_{\rm min}$, whereas the third
chain is still free (intermediate configuration). Note that the
molecule does fluctuate between the two conformations, as witnessed
both by the bimodal character of the probability distribution and
by the non-vanishing values of $P(E_t)$ for
$-\varepsilon < E_t < 0$. Upon further lowering of the
temperature at $T^* = 0.1$, Fig.\ \ref{fig1:plot}(c),
$P(E_t)$ develops a peak
at $E_t = -3\varepsilon$, corresponding to a configuration
in which {\it all three} terminal monomers are confined at 
a distance $r_{\rm min}$ from each other. This is the state in which the micelle has aggregation points at both ends, which we termed `water melon' ({\it wm})-conformation. 
The trend described above
persists for $N = 50$.
However,  
in each panel 
we can notice a higher probability to have association among chains
by decreasing $N$.
We further found a considerable sharpening of the peak at $E_t=-3\varepsilon$
for lowered temperatures, since fluctuations around the ground state are 
getting more suppressed. 

Similar results have been found for the
cases $f=2$, $4$, and $5$. At fixed $N$, the probability of chain
association grows increasing $f$.
The minimum value
of $E_t$ for $f = 2$ is $-\varepsilon$ and for $f=4$ it is
$-6\varepsilon$, corresponding to a {\it wm}-configuration in which
{\it every} end-monomer is confined at a distance $r_{\rm min}$
from every other one, i.e., an arrangement at the 
vertices
of a regular tetrahedron of edge length $r_{\rm min}$. Due to
geometrical constraints, for $f = 5$ it is not possible to have
all five end-monomers at contact with each other; the minimum
value for $E_t$ is $-9\varepsilon$ in this case, 
corresponding to the arrangement of the terminal segments on
the vertices of {\it two} regular tetrahedra of edge length
$r_{\rm min}$, which share a common face.

\begin{figure}
\begin{center}
\includegraphics[width=14.cm,angle=0.,clip]{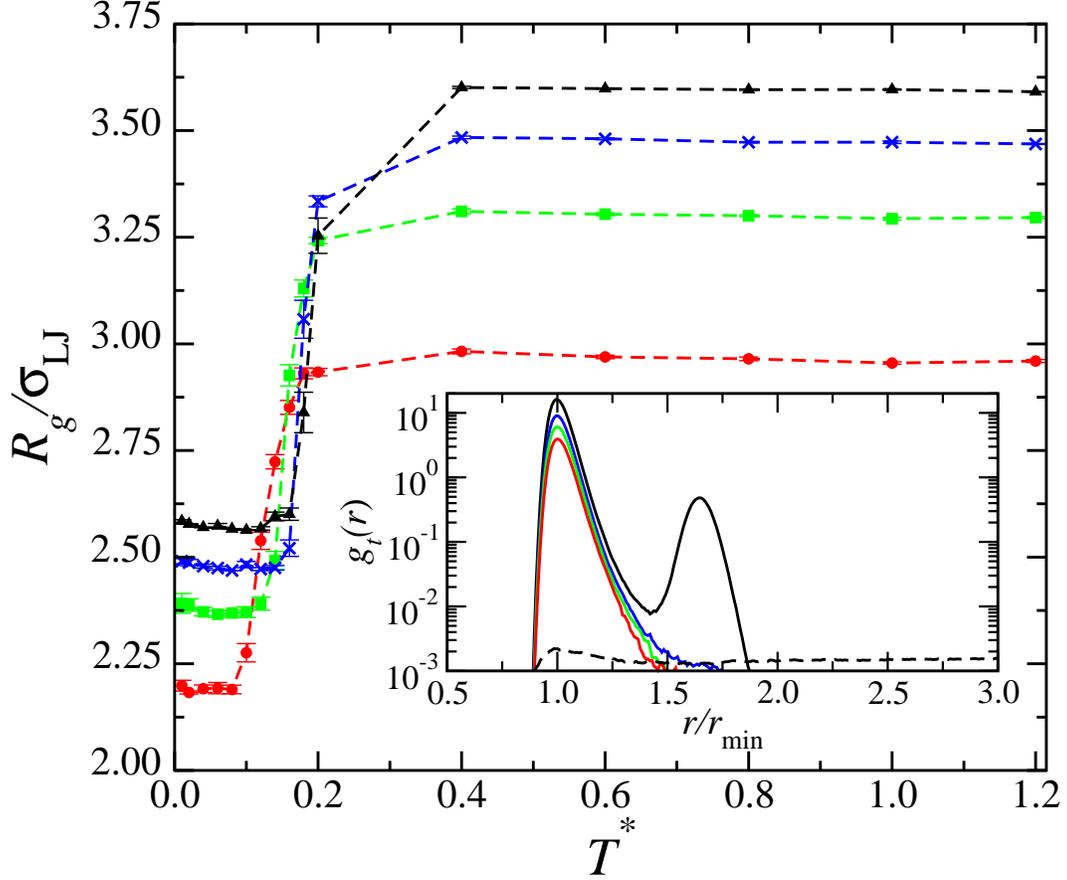}
\caption{(Color online) Temperature dependence of the gyration radius $R_g$
obtained by MD 
of telechelic micelles with
$N=10$. From top to bottom: $f = 5$ (black), $f = 4$ (blue),
$f = 3$ (green), and $f = 2$ (red). Inset: the end-monomer radial
distribution function $g_t(r)$ obtained from MD at $T^* = 0.1$
(solid curves); the sequence from top to bottom and the color coding
are the same as those in the main plot. The dashed line is
$g_t(r)$ for $f = 3$ and $T^* = 1.0$ for comparison.
For clarity, each curve has been multiplied by a different
constant.}
\label{fig2:plot}
\end{center}
\end{figure}
The temperature dependence of the gyration radius $R_g$ for
fixed $N = 10$ and different $f$-values is shown in the main plot of
Fig.\ \ref{fig2:plot}. At high temperatures, $R_g$ has a plateau
that corresponds to {\it sb}-case and scales as $f^{1/5}N^{3/5}$
\cite{daoud:cotton:82}. 
As $T^*$ is lowered, we find a rapid decrease of $R_g$ within
a narrow temperature range and a saturation to a lower plateau
value that corresponds to the size of the {\it wm}-configuration.
The attachment of the terminal monomers leads to a shrinking
of the molecule, in agreement with the experimental findings
of collapsed soft-sphere conformations \cite{vlassopoulos:jcp:99}.
The {\it wm}-configuration persists to higher $T^*$-values upon
increasing the functionality $f$, a trend that can be attributed to
the increasingly strong attractive-energy contributions to the
{\it wm}-free energy as $f$ grows. In the inset of Fig.\ \ref{fig2:plot}
we show the terminal-segment radial distribution function for 
the various $f$-values at $T^* = 0.1$, where the {\it wm}-conformation
is stable and compare it with the one obtained for $f=3$ at 
$T^* = 1.0$, where the micelles assume a {\it sb}-conformation.
The aggregation of the end-monomers is clearly witnessed by the
high peak at $r = r_{\rm min}$, which is present at $T^* = 0.1$,
as opposed by the flat shape of $g_t(r)$ at $T^* = 1.0$.
Whereas for $f = 2$, $3$ and $4$ a single accumulation peak can
be seen, a double peak is present for the case $f=5$. This feature
arises from the arrangement of the five terminal monomers at the
vertices of two tetrahedra that share a common face. The
most distant vertices of the two are separated by a distance
$\sqrt{8/3}r_{\rm min}$, at which indeed the second peak in 
$g_t(r)$ shows up. 

%
%
%
\begin{figure}
\begin{center}
\includegraphics[width=14.cm,angle=0.,clip]{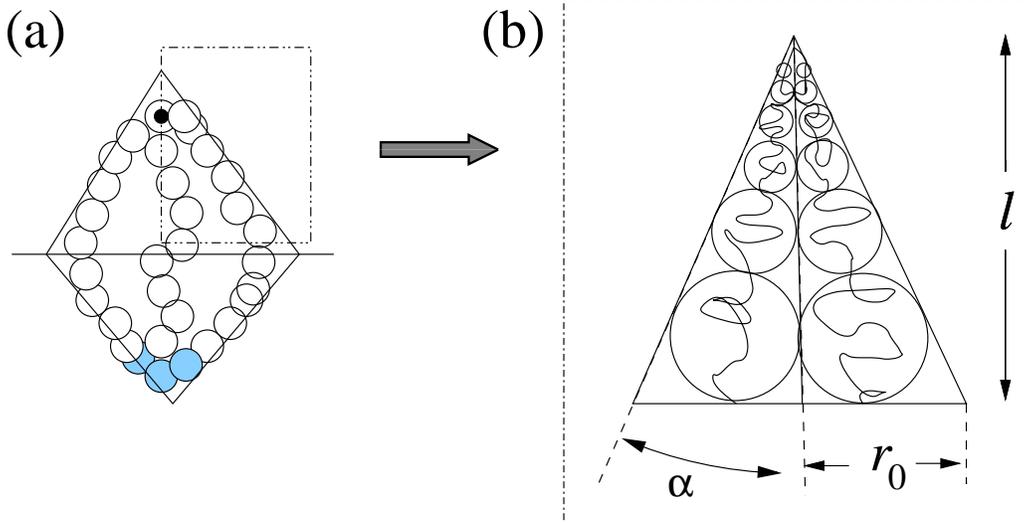}
\caption{(a) Schematic representation of a water-melon with $f=3$ arms,
modeled as a double cone of half-chains;
(b) the corresponding blob model of two arbitrary, attached arms.}
\label{fig3:plot}
\end{center}
\end{figure}
For the theoretical analysis, we consider the 
{\it sb}- and {\it wm}-configurations and put forward 
a scaling theory, to determine the free energy $F$ of each and find the
stable one (lowest $F$-value).
The excluded volume parameter $v$ is set to unity,
consistent with the
choice $\sigma_{\rm LJ} = 1$ above ($v \cong \sigma_{\rm LJ}^3$).
In the case of the {\it sb}-configuration, the free energy is
entirely entropic in nature and can be expressed as a sum of the
elastic and excluded-volume contributions. For a single
chain, minimization of this sum with respect to the 
chain radius $R_g$ leads to the scaling laws $R_g \sim N^{3/5}$ 
and $F \sim k_{\rm B}TN^{1/5}$. For high-functionality star 
polymers, Flory theory can be improved in order to better take
into account the inter-chain correlations by invoking the blob
model of Daoud and 
Cotton \cite{daoud:cotton:82, witten:pincus:86}. For the small
$f$-values at hand, we envision instead each chain as a blob of
radius $\sim R_g$ that occupies a section of space delimited by
a solid angle $4\pi/f$. Accordingly, the total free energy of the
{\it sb}-configuration is approximated by 
$F_{sb} = k_{\rm B}TfN^{1/5}$. The attractive
contribution 
of the terminal monomers is very small, since the latter are far
apart in the {\it sb}-state, and it can thus be ignored.

In the {\it wm}-configuration, there are two accumulation points:
the first is the point in which the chains are chemically linked and
the second is at the other end, 
at which the end-monomers stick together. In order to 
properly take into account the monomer correlations within the
macromolecule \cite{grosberg:book},
we employ a blob model to estimate the excluded-volume
contributions to the free energy $F_{wm}$ \cite{zilman:epje:01,degennes:ssp}.
A schematic representation of the 
water melon and the associated blob model are
shown in Fig.\ \ref{fig3:plot}; the {\it wm} is thereby modeled
as a double cone. With $r_0$ and $l$ being the radius and height
of the cone that contains $n_b$ blobs of a single chain, we obtain
\begin{equation}
\frac{F_{wm}}{k_{\rm B}T}
=\frac{3}{2}f\frac{r_0^2+l^2}{N}+2fn_b+E_{\rm attr}.
\label{Fwm:eq}
\end{equation}
The first term at the right-hand side of Eq.\ (\ref{Fwm:eq}) above
is the stretching contribution of $f$ chains, each having an extension
$\sqrt{r_0^2+l^2}$. The second is the excluded-volume
cost and arises from the total number of $2fn_b$ blobs,
each contributing an amount $k_{\rm B}T$
\cite{zilman:epje:01,grosberg:book,degennes:ssp}.
Finally the third term is the attractive energy contribution arising
from the total number of contacts between terminal monomers.
Accordingly, $E_{\rm attr} = -f(f-1)/(2T^*)$ for $f = 2$, $3$, and $4$,
whereas $E_{\rm attr} = -9/T^*$ for $f = 5$.
The blobs are close-packed within each
cone, hence:
\begin{equation}
\frac{l}{\cos(\alpha/2)}=D\sum_{n=0}^{n_b-1} 
\left(\frac{1+\sin(\alpha/2)}{1-\sin(\alpha/2)}\right)^{n},
\label{constr1:eq}
\end{equation}
where $\alpha=\arctan(r_0/l)$ is the cone opening angle 
and $D$ is the diameter of the first, 
smallest blob,
taken equal to the monomer size.
The polymerization $N$ is obtained by summing over the 
monomers of all blobs of a chain, taking into account that  
a blob of size $b$ contains $N_b \sim b^{5/3}$ monomers. This 
yields
\begin{equation}
N(r_0, l)=2D^{5/3}\sum_{n=0}^{n_b-1} 
\left(\frac{1+\sin(\alpha/2)}{1-\sin(\alpha/2)}\right)^{(5n/3)}. 
\label{constr2:eq}
\end{equation}
We solved Eqs.\ (\ref{constr1:eq}) and (\ref{constr2:eq}) numerically to
express $n_b$ and $l$ in terms of $r_0$ and then we minimized
numerically $F_{wm}$ with respect to $r_0$ using Eq.\ (\ref{Fwm:eq}).
We found that the cone angle $\alpha$ between two arms
increases rapidly with $N$ for small 
$N$-values: for $5\lesssim N \lesssim 20$
it changes from $20$ to $40$ degrees. 
This trend reflects the physical difficulty
for the chains to
form an aggregate
for short chains, due to strong steric hindrance at the 
star center.
Once the excluded volume
interactions are balanced, 
the angle increases slowly, by about $5$ degrees 
for $25\lesssim N \lesssim 400$. 
\begin{figure}
\begin{center}
\includegraphics[width=14.cm,angle=0.,clip]{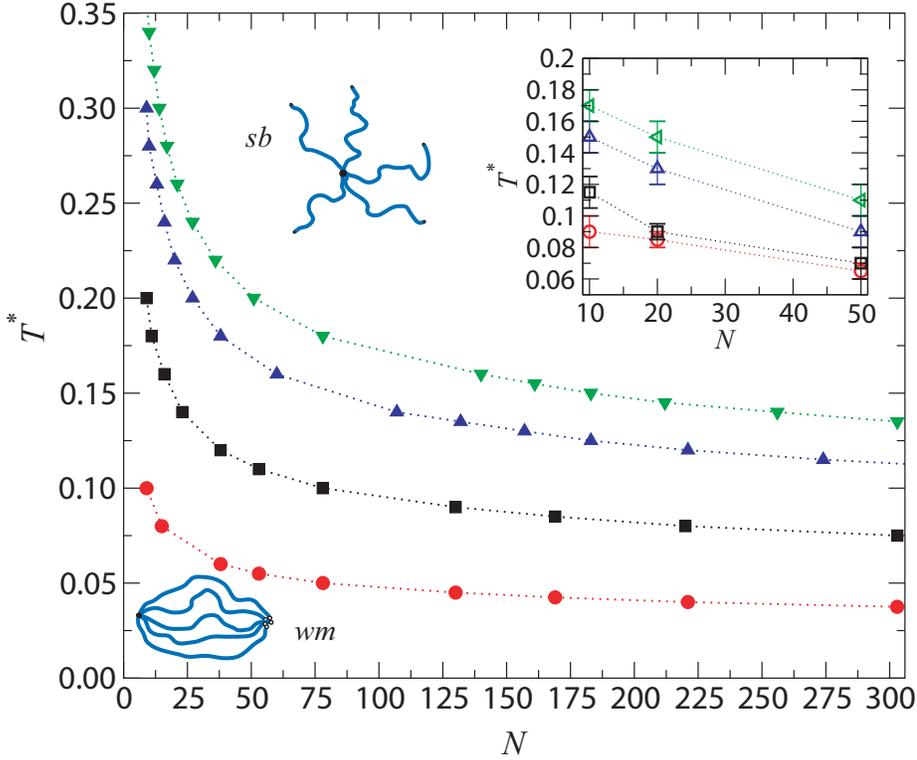}
\caption{Temperature versus $N$ diagram corresponding to 
the {\it sb}- and {\it wm}-morphological transitions
of telechelic micelles. Main figure: theoretical results.
Inset: transition lines in simulations, determined 
via the $T^*$ dependence of $R_g$ .
From bottom to top  (full/empty symbols): $f=2$ (circle), $f=3$ (square), $f=4$ (triangle up), $f=5$ (triangle down).}
\label{fig4:plot}
\end{center}
\end{figure}

Comparing the minimized free energies $F_{wm}$ 
with 
$F_{sb}$, 
we obtained
the {\it transition} temperature 
between $sb$  and {\it wm}  as a function of
$N$ \footnote{As we are dealing with finite systems,
these are not sharp transitions but rather crossover phenomena.}.
In Fig.\ \ref{fig4:plot} we show the $T$ vs.\ $N$ transition lines,
separating the {\it wm}-state (below) from the {\it sp}-one (above). 
The `critical number' $N_c$ increases  with decreasing $T^*$
and the transition temperature 
increases with $f$.
Stated otherwise, 
for fixed $N$ the theoretical model gives evidence to a stronger
stability of the {\it wm} configuration on
increasing $f$ and lowering the temperature.
On the other hand, keeping 
$f$ and $T^*$
fixed,
the {\it sb}-configuration is stable above a certain value of $N_c$.
The dependence of the repulsive contribution in the free energy 
on the number of monomers is not trivial:
increasing $N$, the number of blobs $n_b$ increases and so does
the contribution due to the excluded volume interaction.
The entropic contribution is proportional to 
$r_0^2$, which increases with $N$, and to $1/N$.
The balance between all these terms in Eq.\ (\ref{Fwm:eq}) 
determines the stable micelle conformation:
it is physically more expensive 
for the molecule 
to assume a {\it wm}-configuration 
the lower $f$ and the higher $N$ is. 
$N_c$ decreases rapidly with $T^*$ and the {\it wm}-state becomes
unstable at $T^* \gtrsim 0.3$ for all $f$-values considered.
In the inset of Fig.\ \ref{fig4:plot} we
show the corresponding MD-results:
notice the nice agreement with the theoretical prediction trends.
The value of transition temperature is overestimated in theory,
reflecting the mean-field nature of the latter.
There exist also configurations
intermediate to the {\it wm}- and {\it sb}-ones, in which only
a partial association of arms takes place,
akin to those seen for block-copolymer stars with $f=12$
in the simulation study of Ref.\ \cite{connolly:jcp:03}. Such configurations
are also expected to be dominant as $f$ further increases, due
to the very high entropic penalty inherent to a complete
intramolecular association. These conformations can be analyzed
along the lines put forward in this work, since
intermediate states can be looked upon as a combination
of smaller water melons (work in progress).

%
%
%
%
We have studied the dependence of the equilibrium
conformations of telechelic micelles
on the temperature, arm number and functionality of the same.
The system presents a stable, water-melon configuration 
for low temperatures that opens up as $T$ grows.
The dynamical
properties
of such aggregates in solution will have unusual
characteristics: at low temperatures and near
the overlap concentration, various types of gels can appear, 
owing their viscosity either to transient network formation or
to the existence of entangled loops.
An understanding of the changes in  
the micelle conformations is a relevant starting point, aiming at
gaining control over the 
supramolecular structural and dynamical properties of the 
system \cite{thanassis:prl:05}, 
and targeting specific applications.

We thank W.\ Russel for bringing Refs.\ \cite{russel:macrom:03}
and \cite{russel:macrom:00} to our attention and 
D.\ Vlassopoulos for helpful discussions.
C.M.\ thanks the D\"usseldorf Entrepeneurs Foundation for 
financial support.
This work was supported by the DFG within the SFB-TR6 and by the 
Marie Curie European Network MRTN-CT-2003-504712.


\end{document}